\documentclass{article}

\usepackage[dvipsnames]{xcolor}

\definecolor{myblue}{RGB}{31, 119, 180}
\definecolor{myorange}{RGB}{255, 127, 14}
\definecolor{mygreen}{RGB}{44, 160, 44}
\definecolor{myred}{RGB}{214, 39, 40}

\usepackage{natbib}
\usepackage{thmtools} 
\usepackage{thm-restate}
\usepackage{lipsum}
\usepackage{wrapfig}
\usepackage{makecell}
\usepackage[normalem]{ulem}
\usepackage[margin=1in]{geometry}
\usepackage[utf8]{inputenc} %
\usepackage[T1]{fontenc}    %
\usepackage[allcolors=NavyBlue,colorlinks=true,backref=page]{hyperref}       %
\usepackage{url}            %
\usepackage{tabularx}
\usepackage{booktabs}       %
\usepackage{amsfonts}       %
\usepackage{nicefrac}       %
\usepackage{microtype}      %
\usepackage{xspace}

\usepackage{authblk}

\usepackage{xcolor}

\usepackage{longtable}
\usepackage{caption}
\usepackage{array}
\usepackage{multirow}
\usepackage{color, colortbl}
\definecolor{lightGray}{gray}{0.8}

\usepackage{amsmath}
\usepackage{textcomp}
\newcommand*\mean[1]{\bar{#1}}

\usepackage[labelformat=simple,skip=3pt]{subcaption}

\newcommand{\xhdr}[1]{\vspace{1.3mm}\noindent{{\bf #1.}}}

\usepackage{tabularray}

\usepackage{wrapfig}

\usepackage[textsize=scriptsize]{todonotes}

\newif\iffinal

\iffinal
    \newcommand{\fix}[1]{}
    \newcommand{\XL}[1]{}
    \newcommand{\XLinline}[1]{}
    
    \newcommand{\SJnote}[1]{}
    \newcommand{\SJ}[1]{}
    \newcommand{\note}[1]{}
    \newcommand{\pref}[1]{}
    \newcommand{\hm}[1]{}
    \newcommand{\modify}[1]{{}}
\else
    \newcommand{\modify}[1]{{\color{green}\sout{#1}}}
    \newcommand{\fix}[1]{{\color{red} #1}}
    \newcommand{\XL}[1]{\todo[fancyline,color=red]{XL: #1}\xspace}
    \newcommand{\SJ}[1]{\todo[fancyline,color=blue]{SJ: #1}\xspace}
    \newcommand{\XLinline}[1]{\textcolor{red}{[XL: #1]}}
    
    \newcommand{\SJnote}[1]{\textcolor{green}{[SJ: #1]}}
    
    \newcommand{\note}[1]{{\color{purple}[XL: #1]}}

    \newcommand{\pref}[1]{{\color{blue}(\ref{#1})}}
    \newcommand{\hm}[1]{{\textcolor{purple}{[**Heng: #1 ]}}\xspace}
\fi

\newcommand{\figref}[1]{Fig.~\ref{#1}}

\title{Binding Affinity Prediction: From Conventional to Machine Learning-Based Approaches}

\author[1,2,$*$]{Xuefeng Liu}
\author[1]{Songhao Jiang}
\author[1,2]{Xiaotian Duan}
\author[2]{Archit Vasan}
\author[3]{Qinan Huang}
\author[4,5]{Chong Liu}
\author[6]{Michelle M.~Li}
\author[2]{Heng Ma}
\author[2]{Thomas Brettin}
\author[2]{Arvind Ramanathan}
\author[2]{Fangfang Xia}
\author[7]{Mengdi Wang}
\author[8]{Abhishek Pandey}
\author[6]{Marinka Zitnik}
\author[1,2]{Ian T.~Foster}
\author[9,$*$]{Jinbo Xu}
\author[1,2,$*$]{Rick L.~Stevens}

\affil[1]{\small Department of Computer Science, University of Chicago}
\affil[2]{Argonne National Laboratory}
\affil[3]{Pritzker School of Molecular Engineering, University of Chicago}
\affil[4]{Data Science Institute, University of Chicago}
\affil[5]{Department of Computer Science, State University of New York at Albany}
\affil[6]{Department of Biomedical Informatics, Harvard Medical School}
\affil[7]{AI Lab, Princeton University}
\affil[8]{Information Research, AbbVie Inc.}
\affil[9]{Toyota Technological Institute at Chicago}
\affil[$*$]{Correspondence: \href{mailto:xuefeng@uchicago.edu}{xuefeng@uchicago.edu}, \href{mailto:jinboxu@gmail.com}{jinboxu@gmail.com}, \href{mailto:rstevens@uchicago.edu}{rstevens@uchicago.edu}}

\date{}

\begin{document}
\maketitle

\begin{abstract}
Protein-ligand binding is the process by which a small molecule (drug or inhibitor) attaches to a target protein. Binding affinity, which characterizes the strength of biomolecular interactions, is essential for tackling diverse challenges in life sciences, including therapeutic design, protein engineering, enzyme optimization, and elucidating biological mechanisms. Much work has been devoted to predicting binding affinity over the past decades. Here, we review recent significant works, with a focus on methods, evaluation strategies, and benchmark datasets. We note growing use of both traditional machine learning and deep learning models for predicting binding affinity, accompanied by an increasing amount of data on proteins and small drug-like molecules. 
With improved predictive performance and the FDA's phasing out of animal testing, AI-driven \textit{in silico} models, such as AI virtual cells (AIVCs), are poised to advance binding affinity prediction; reciprocally, progress in building binding affinity predictors can refine AIVCs. Future efforts in binding affinity prediction and AI-driven \textit{in silico} models can enhance the simulation of temporal dynamics, cell-type specificity, and multi-omics integration to support more accurate and personalized outcomes.

\end{abstract}

\section{Introduction}

\begin{figure}
        \centering
\includegraphics[width=1\linewidth]{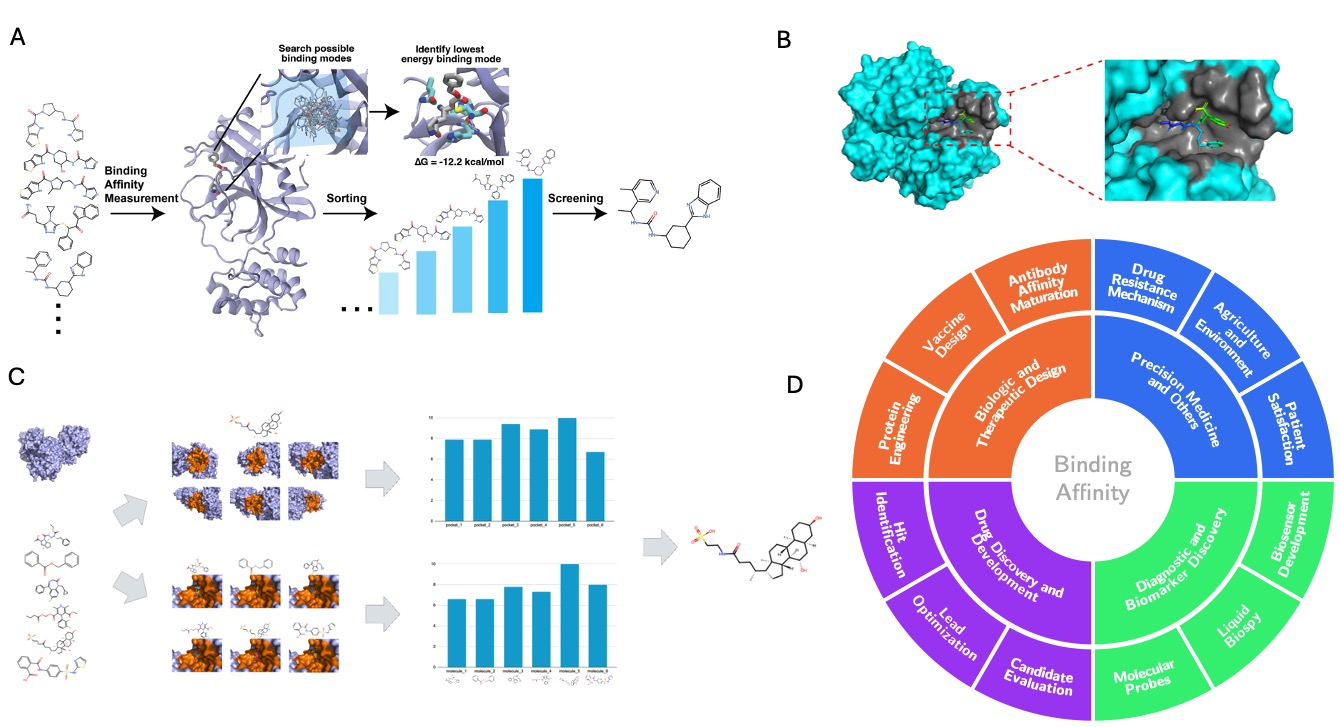}
        \caption{Overview of binding affinity. \textbf{(A) Binding affinity pipeline.} Computational binding affinity measurements are typically performed using molecular docking simulations as a surrogate. For each compound, the simulations search for an optimal binding pose and produces a score. These scores are then used to rank compounds with respect to each other.\textbf{(B) Binding Example in Surface View with Protein PDB: 10GS.}  \textbf{(C) Detailed pipline.} For a given protein and a set of candidate molecules, we perform both single-site docking and multi-site docking. The top row illustrates the search for optimal molecules and corresponding binding poses across multiple potential binding pockets, while the bottom row depicts the docking of various molecules into a single predefined pocket. By combining these two strategies, we aim to identify the most favorable molecule, binding pose, and docking score.
        \textbf{(D) Application domains.} Binding affinity is crucial across drug discovery, biologics design, diagnostics, and precision medicine. It guides the identification and optimization of molecules—such as small drugs, antibodies, or probes—for strong and selective target binding.
        }
        \label{fig:enter-label}
    \end{figure}

    Protein-ligand binding~\citep{clyde2023ai} refers to the process by which ligands—usually small molecules, ions, or proteins—generate signals by binding to the active sites of target proteins through intermolecular forces (\figref{fig:enter-label}). This process typically changes the conformation of target proteins, which then results in the realization, modulation, or alteration of protein functions. As such, 
    protein–ligand binding affinity is a fundamental parameter in understanding and optimizing molecular interactions, with broad applications across biomedical research and drug development. In small-molecule drug discovery, it guides hit identification, lead optimization, and candidate selection to ensure strong, selective binding to target proteins. In biologics and therapeutic protein design, it informs the engineering of enzymes, receptors, and antibodies for improved specificity and potency. In diagnostics, it underlies the development of high-sensitivity biosensors and molecular probes for detecting disease biomarkers. In precision medicine, protein–ligand affinity data help predict patient-specific drug responses, uncover mechanisms of resistance, and tailor treatments, while also finding utility in agricultural and environmental applications through the design of selective binding agents.
    
    Binding affinity prediction was introduced in \citet{bhm1994}: Given the 3D structures of a target protein and a potential ligand, predict the binding constant of the complex along with the most probable binding pose candidates. The prediction of the binding site (i.e.,~the set of protein residues that have at least one non-hydrogen atom within 4.0 \r{A} of a ligand's non-hydrogen atom \citep{khazanov2013}) and affinity (i.e.,~binding constants such as inhibition or dissociation constants, or the concentration at 50\% inhibition) are usually divided into two separate but related stages  \citep{ballester2010}.

        One notable motivation for constructing a good binding affinity predictor (or scoring function) is the essential role that it plays in drug discovery~\citep{liu2023drugimprover, liu2024entropy} and virtual screening~\citep{meng2011,pinzi2019molecular, sadybekov2023computational}. Traditional drug discovery involves a process of trial and error; however, with a functional binding affinity predictor, we can significantly reduce the number of experimental trials by focusing only on drugs estimated to be more effective, thereby reducing costs~\citep{scannell2012,hay2014,tiwari2023artificial}. %
        Beyond its economic significance in drug development, binding affinity prediction also provides tools, methods, and insights into many other areas of research in biochemistry, pharmaceutical research, and scientific computation.

    Despite the increasing motivation and extensive amount of work, binding affinity prediction remains challenging. First, our understanding of chemistry is not comprehensive, %
    which often leads to sub-optimal human-engineered features for binding affinity prediction~\citep{ballester2010, ballester2014}. Secondly, synthetic datasets remain undesirable for learning because computationally generated complexes are either inaccurate or too costly to generate in terms of time and computational resources.
    The data sets currently being used, which are usually obtained experimentally, are limited in various ways: (1) the total number of data points is increasing but still far from sufficient for large-scale data mining due to the vast time and money required to study each complex; (2) the precision of the experimental data may be limited by the measurement methods used~\citep{su2018, pantsar2018}; and (3) samples are often biased toward complexes that have the correct poses and good binding constants, meaning that protein-ligand pairs that do not bind or have relatively low binding constants are not readily obtainable. %
    These limitations make the search for a good protein-ligand binding affinity predictor extremely challenging.
    
    Further, there are different aspects/sub-problems associated with the prediction of binding constants.  \citet{liyan2014_2} and \citet{liu2017} note  four different types of predictions related to binding affinity: (1)~scoring, which predicts the binding constant; (2)~rank ordering, which ranks different potential ligands of a target; (3)~docking, which predicts the best binding pose from multiple decoys; and (4)~screening, which predicts the best ligand from multiple decoys. These sub-problems are closely related to each other, to the extent that solving one without addressing the others adds to the existing challenges. Thus, despite existing efforts to improve protein-ligand binding affinity prediction, the results remain sub-optimal. It is often the case that a predictor excels in scoring for known targets but fails miserably in docking or screening for new targets, rendering the whole system almost useless for virtual screening, one of the ultimate goals of binding affinity prediction.

    Existing methods for binding affinity prediction can be roughly divided into three distinct categories: conventional, traditional machine-learning-based, and deep-learning-based. Conventional methods are typically based on ab initio quantum mechanical calculations or empirical approaches derived from experimental data, often formulated as physics-based models or parametric equations that predict binding affinity(\figref{fig:enter-label}(A)). These methods are usually rigid and tend to work well only in specific scenarios, such as with a single protein family~\citep{bender2021practical}.
    Since 2005, many traditional machine learning methods %
    have been applied to human-engineered features extracted from complex structures, achieving some improvement over conventional approaches. Machine learning methods have proven to be less rigid and often more accurate, especially in terms of binding affinity scoring and ranking \citep{ballester2010machine, zilian2013, li2011}
    A decade later, as the number of protein-ligand samples in standard benchmarks like PDBbind~\citep{wang2004pdbbind}, MOAD~\citep{hu2005}, and PDB~\citep{burley2017protein} increased, deep learning began to dominate. This approach typically relies less on human-engineered features, if at all, and its learning potential could greatly increase with the continued accumulation of data.

    Looking ahead, the increasing emphasis on in silico drug discovery—accelerated by the FDA’s move to phase out animal testing—positions AI-driven computational models as a transformative force in binding affinity prediction. In particular, emerging AI virtual cells (AIVCs) offer a systems-level framework for modeling molecular interactions in dynamic, cell-specific, and multi-omic contexts. Advances in binding affinity predictors will not only strengthen the molecular foundations of AIVCs but also benefit from the broader simulation capacities that AIVCs provide. This reciprocal relationship underscores a future in which binding affinity prediction and AI-driven in silico modeling co-evolve to enable more accurate, mechanistic, and personalized biomedical insights.

    This Review covers foundational research on protein-ligand binding affinity prediction, from the early 2000s to the present, along with commonly used datasets and evaluation metrics. It is organized as follows. \autoref{sec:2} introduces important datasets and benchmarks, including data specifications and various datasets. 
    \autoref{sec:3} discusses conventional and machine learning-based methods. 
    \autoref{sec:4} introduces various evaluation approaches, including scoring power, docking power, and ranking power. Finally, we discuss open questions and research directions that remain unexplored in the field.

\section{Datasets and Benchmarks} %
\label{sec:2}

    There are many datasets and benchmarks used for the study of protein-ligand binding affinity prediction, each focusing on different aspects of the problem. In \autoref{subsec:db_ds}, we discuss the specifications for binding affinity data, such as resolution, measurement, binding constant, and concentration. These factors are essential when selecting the right dataset for a specific goal. In \autoref{subsec:db_ba}, we review some of the most commonly used datasets and benchmarks for developing models to predict protein-ligand binding affinity (\autoref{tab:datasets}).
    These datasets usually have binding constants such as $K_i$, $K_d$, or $IC_{50}$ associated with most, if not all, protein-ligand complexes.
    We also note 
    datasets of protein-ligand binding affinity not directly used for model training and evaluation. While these datasets are not directly applicable for training and evaluation, they can be useful for pre-training and may provide additional information, methods, and insights.

\begin{center}
\begin{table}

    \centering
    \renewcommand*{\arraystretch}{1.5}
        \scalebox{0.75}{
    \begin{tabular}{cccccc>{\centering\arraybackslash}m{4cm}}
        \toprule
        Name & 
            \# Proteins & 
            \# Ligands & 
            \# Complexes & 
            \# Affinities &
            3D& 
            Primary Sources \\
        \midrule
        \rowcolor{lightGray}
        PDBbind  & 
            9,198 & 
            13,427 & 
            19,588 &
            19,588 &
            Yes & 
            PDB \\
        CASF  & 
            57 & 
            285 & 
            285 &
            285 &
            Yes & 
            PDB \\
        \rowcolor{lightGray}
        Binding MOAD & 
            9,117 & 
            16,044 & 
            32,747 &
            12,101 &
            Yes & 
            PDB \\
        BindingDB & 
            7,317 & 
            751,447 & 
            1,692,135 & 
            1,692,135 & 
            Partial & 
            Publications, PubChem, ChEMBL\\
       \rowcolor{lightGray}
       DUD-E & 
            102 & 
            22,886 & 
            22,886 & 
            22,886 & 
            Yes & 
            PDB, ChEMBL, ZINC \\
        BioLiP &
            97,966 &
            460,356 &
            460,364 &
            23,492 &
            Yes &
            Binding MOAD, PDBbind, BindingDB, Pubications \\
        \rowcolor{lightGray}
        PDSP Ki &
            958 & 
            12,228 & 
            67,689 & 
            67,689 & 
            No & 
            Publications \\
        KIBA & 
            467 & 
            52,498 & 
            246,088 & 
            246,088 &
            No & 
            ChEMBL  \\
            
        \bottomrule
    \end{tabular}}
    \caption{Commonly Used Protein-Ligand Binding Affinity Datasets}
    \label{tab:datasets}
\end{table}\end{center}

    \subsection{Data Specifications} \label{subsec:db_ds}
    
    \subsubsection{Structure}
    
        The structure of a protein-ligand complex refers to its conformation (or the spatial arrangement of its atoms). This is an important specification or feature of binding affinity data, as it opens up numerous possibilities for feature engineering and learning methods.
        For instance, with given structures, we can parameterize the energy terms and atom-pair distances and use them for conventional scoring or learning with methods such as random forests \citep{ballester2010}, support vector machines \citep{kinnings2011}, and neural networks \citep{durrant2010, durrant2011}. However, measuring structures is rather expensive (more than \$1,000 USD for solving a structure by crystal x-ray diffraction) compared to sequence identity, along with the additional costs of data storage and manipulation. Still, the advantages of having a well-measured structure significantly outweigh the disadvantages, and we are observing a rising trend of more protein-ligand complexes with structures being added to the Protein Data Bank (PDB) and other databases.
        
    \subsubsection{Resolution}
        Resolution, in the context of protein-ligand complex structures, refers to the distance corresponding to the smallest observable feature in the measured structure \citep{guterres2020improving}. In other words, two objects within this distance will be perceived as one and thus rendered indistinguishable. 
        Generally, structures with a resolution finer than 1--1.2 \AA~are considered high-resolution, while those with a resolution lower than 3.0 \AA\ only outline the basic contours of the protein chains. Resolution is an essential measurement for the quality of data and is used as a selection criterion for many binding affinity databases (e.g.,~CASF, Binding MOAD). Importantly, the effective resolution at the binding site is often more critical than the global resolution of the structure, since small errors in ligand–protein contact regions can disproportionately affect the accuracy of binding affinity estimation and downstream modeling.

    \subsubsection{Structure determination techniques}

        The three most commonly used techniques for determining the structure of protein complexes are X-ray crystallography~\citep{maveyraud2020protein,jackson2023general}, NMR (nuclear magnetic resonance)~\citep{hore2015nuclear,galvan2023successful}, and Cryo-EM (cryo-electron microscopy)~\citep{adrian1984cryo,nogales2024bridging}.
        
        Roughly speaking, X-ray crystallography determines the position and arrangement of atoms in a single crystal of the target protein by examining the diffraction intensity obtained with X-rays. 
        Given high-quality protein crystal, X-ray crystallography is capable of generating atom-level static structures, regardless of molecular weight of the samples. 
        
        NMR is a fundamentally different technique for protein structure determination. The analysis is performed on a solution of the target protein with high purity and high concentration to obtain characteristic NMR signals, which are interpreted by computer-aided methods to determine the 3D structures. 
        The most notable feature of NMR is that it allows us to obtain the dynamic structure of the target protein in its natural state in solution (without crystallization). However, the measurement is complicated, requires computational interpretation (making NMR an indirect method of structure determination), and is not applicable to large molecules or samples without pure and highly-concentrated solutions.
        
        Cryo-EM uses a mechanism called electron scattering, where electron beams pass through an instantly cooled protein solution, scatter into a lens, and are converted into a series of 2D images on a detector. These images are processed by reconstruction software to obtain the 3D structure. Cryo-EM is becoming increasingly popular because it works well with larger proteins, has an easier sample preparation process, and yields 3D structures that are much closer to their native state than those obtained from X-ray crystallography. However, the resolution of the structures can be relatively low due to the high level of noise and unknown orientations.

    \subsubsection{Binding affinity measurement}
        
        Binding affinity measures how ``tightly'' a ligand binds to a target protein, which is determined with radioligand-binding experiments \citep{haylett2003direct}. 
        A radioligand-binding experiment can produce two kinds of binding affinity data: (1) binding constants, such as dissociation constant ($K_d$) and inhibition constant ($K_i$), which are usually interchangeable in the context of protein-ligand binding; and (2) concentration terms, like the half maximal inhibitory concentration (${IC}_{50}$) or half maximal effective concentration (${EC}_{50}$).
        In binding affinity databases, each protein-ligand complex is typically associated with one or more binding data of $K_d$, $K_i$, and ${IC}_{50}$. %
        In certain databases and studies, like the \textit{refined set} in PDBbind, complexes with concentration terms only will be filtered out for better data quality.
        This is due to the fact that ${IC}_{50}$ or ${EC}_{50}$ values depend on the radioligand concentration and can vary between different experiments, unlike constant terms. 
        If necessary or useful, the inhibitory constant \(K_i\) value can be estimated from \(IC_{50}\) via the Cheng-Prusoff equations:
        \begin{equation*}
            K_i = \frac{\text{IC}_{50}}{\frac{[A]}{\text{EC}_{50}}+1} \;\text{,}
        \end{equation*}
        where \([A]\) is a fixed concentration of the ligand.  However, the estimation is prone to inaccuracy with high or low concentration values \citep{lazareno1993estimation}.

    \subsection{Binding Affinity Datasets} \label{subsec:db_ba}

    \xhdr{PDBbind}
        Starting in 2004, PDBbind has been updated annually \citep{wang2005, liu2015, liu2017}. It collects not only protein-ligand complexes but also protein-protein, protein-nucleic acid, and nucleic acid-ligand complexes from the Protein Data Bank (PDB) \citep{berman2000} without any restrictions on resolution, binding data, or structural measurement techniques.
        An important feature of PDBbind is the \textit{refined set}, selected based on (1) the quality of the complex structure, (2) the quality of binding data, and (3) the nature of the complex (e.g.,~molecular weight, atom types, surface area)~\citep{liu2017}. %
        However, the \textit{refined set}, despite its much higher quality compared to the \textit{general set}, should not be regarded as a high-quality dataset but rather as a collection of complex samples that lack any obvious problems~\citep{liu2017}.
        With the increasing number of data points, subsets with quality control, and easy accessibility, PDBbind is becoming one of the most used datasets in the research of protein-ligand binding affinity.%
    
    \xhdr{CASF (Comparative Assessment of Scoring Functions)}
        CASF \citep{cheng2009, li2018, su2018} was introduced along with PDBbind.
        Originally named the \textit{core set }in PDBbind, CASF was selected from the \textit{refined set} and serves as the test set or benchmark for protein-ligand binding predictors. 
        To ensure that the samples in the test set are diverse but not redundant, all complexes in the \textit{refined set} are first clustered based on sequence similarity. Then, for each cluster, three complexes, each with high, medium, and low binding affinity, are selected for inclusion in CASF.
        With this type of data selection, we can not only test regression on binding affinity (scoring) but also ranking (three different complexes in the same cluster). Additionally, with the given decoys of poses and ligands in CASF, we can evaluate the docking and screening power of a predictor.
        Overall, CASF provides researchers with a comprehensive and easy-to-use benchmark for protein-ligand binding affinity predictors.
 
    \xhdr{Binding MOAD (Mother of All Databases)}
        Binding MOAD \citep{hu2005, ahmed2015, smith2019}  is another commonly used dataset for binding studies. Proposed in 2005 and updated annually, Binding MOAD is a subset of PDB, aiming to be the largest collection of high-quality protein-ligand complexes  annotated with experimentally determined binding affinity. 
        Despite sharing the same primary data source (PDB) and a similar goal, Binding MOAD has much in common with PDB.
        The main differences that set these two apart for users are:
        (1) Binding MOAD sets the data selection criteria somewhere between PDBbind's general set and refined set. 
        Notably, Binding MOAD only contains valid protein-ligand complexes with crystal structures of 2.5 \AA resolution or better. 
        Additionally, protein sequences are clustered to avoid redundancy of the data, similar to the refined set in PDBbind.
        (2) Binding MOAD contains complexes without binding data, which is not the case in PDBbind. 
        In fact, Binding MOAD contains only 12,098 binding data of 32,747 complexes.
        This means that Binding MOAD, while having more complexes with structures than PDBbind, contains fewer samples with binding data.

    \xhdr{BindingDB}
        Launched in 2000 and updated weekly as a web-accessible database, BindingDB \citep{liu2007, gilson2016} collects protein-compound pairs with associated affinity data primarily from scientific articles and, increasingly, patents. It also accepts direct deposition of binding data by users. Newly curated data from these original sources are checked by BindingDB staff to ensure reliability. Additionally, it integrates complexes with affinity measurements from PubChem \citep{wang2009pubchem}, ChEMBL \citep{bento2014chembl}, PDSP Ki \citep{roth2000}, and CSAR \citep{carlson2011call}, allowing users to access the data through the unified interface of BindingDB \citep{gilson2016}.
        There are two important differences between BindingDB and other large binding affinity datasets, such as PDBbind and Binding MOAD. First, although most binding data are experimentally determined, BindingDB contains some complexes that are computationally generated with clear labels. 
        Secondly and more importantly, only a fraction of BindingDB's affinity data has associated structures. In fact, according to their website, there are 2,291 protein-ligand crystal structures available in BindingDB with 100\% sequence identity and 5,816 crystal structures with protein sequence identity as low as 85\%. %
        As such, BindingDB is a protein-ligand binding affinity database with the largest number of complexes.

    \xhdr{DUD-E (A Directory of Useful Decoys: Enhanced)}
        To combat data bias and benchmark ligand enrichment against challenging decoys, DUD \citep{huang2006} (Directory of Useful Decoys) and its successor DUD-E \citep{mysinger2012} (A Directory of Useful Decoys: Enhanced) were proposed in 2006 and 2012, respectively. The dataset was constructed by sourcing ligands from ChEMBL, structural data from PDB, and potential decoys from ZINC. The docking poses were generated with DOCK \citep{ewing2001dock, allen2015dock}.
        Immediately after being proposed and constructed, DUD became the gold standard for the evaluation of virtual screening methods \citep{rau2018}. However, as pointed out in \citet{good2008}, the original DUD dataset had a number of issues, such as target selection and analogue bias. To address these issues, the authors of DUD released DUD-E, which added net charge into consideration during decoy selection. In DUD-E, each of the 102 target proteins has 50 decoys that are similar to the ligand in terms of 1D physico-chemical properties, such as molecular weight and calculated LogP, to remove dataset bias, but differ in 2D topology to be likely non-binders.
        Nevertheless, DUD-E has been shown to remain unsatisfactory~\citep{xia2015, chen2019, sieg2019}, indicating that hidden bias in the dataset might still exist, leading to misleadingly good benchmarking results for many docking methods. %

    \xhdr{BioLiP}    
        BioLiP~\citep{yang2013} and its successor BioLiP2 \citep{zhang2024biolip2} (collectively referred to as BioLiP) are semi-manually curated databases for protein-ligand interactions. Founded in 2013 and updated weekly, BioLiP is designed to focus only on biologically relevant ligands, meaning they are not simply additives for protein purification and/or crystallization. This is achieved through automatic procedures and manual verification.
        In addition to annotations of binding affinity, the dataset also contains relevant functional annotations of the complexes. It sources structures from PDB and binding data from multiple databases, such as Binding MOAD, PDBbind, BindingDB, and the literature.
        It is worth noting that the main focus of BioLiP is binding site prediction, similar to FireDB \citep{maietta2014firedb} and LigASite \citep{dessailly2007ligasite}. Binding affinity is one of the optional fields for each entry in the database. As such, only about 5\% of the protein-ligand complexes have associated binding affinity.

    \xhdr{PDSP (Psychoactive Drug Screening Program) Ki Database}
        The PDSP Ki Database \citep{roth2000}, founded in 2006, serves as a regularly updated data warehouse for published binding affinity data of drugs and drug candidates for receptors, ion channels, transporters, and enzymes.
        The entries are accessible for query under given conditions or can be downloaded as a CSV file. For each entry, the IDs of the ligand and receptors are provided, along with additional information, such as the SMILES string and the reference for the affinity data.
        Despite the large quantity of affinity data, there are two major drawbacks to the PDSP Ki Database.
        First, the affinity data ($K_i$ value) for each compound-receptor pair may be either a single numeric value or an inequality, such as ``greater than 7000.'' 
        Secondly, structure features for both protein and ligand are not included in the dataset, meaning users will have to query other data sources such as PDB.
    
    \xhdr{KIBA (Kinase Inhibitor Bioactivity) Dataset}
        KIBA was originally proposed in \citet{tang2014}. It aimed to integrate various bioactivity types, including $K_i$, $K_d$, and $IC_{50}$, into a single term named the KIBA score, and to demonstrate its usage in classifying kinase inhibitor targets and pinpointing errors in binding affinity datasets.
        The dataset was obtained from searching for drug-kinase interactions in ChEMBL with $IC_{50}$ and at least one binding constant terms ($K_i$ and $K_d$). 
        For each interaction in the KIBA dataset, they have converted the bioactivity terms into KIBA score using the equations below: 
        \begin{align}
            {K_i}_{\cdot adj} &= \frac{{IC}_{50}}{1 + L_i ({IC}_{50} / K_i)},~ 
            {K_d}_{\cdot adj} = \frac{{IC}_{50}}{1 + L_d ({IC}_{50} / K_d)} & \\
            \text{KIBA} &= 
                \begin{cases} 
                    {K_i}_{\cdot adj} & \text{if ${IC}_{50}$ and $K_i$ are present} \\
                    {K_d}_{\cdot adj} & \text{if ${IC}_{50}$ and $K_d$ are present} \\
                    ({K_i}_{\cdot adj} + {K_d}_{\cdot adj})/2 & \text{if ${IC}_{50}$, $K_i$, and $K_d$ are present}
                \end{cases}  
        \end{align}
        with $L_i=0.3$ and $L_d=1.3$, which are the optimal values for adjustment. 
        There are no features included; users need to look up ChEMBL, PDB, or other sources to obtain features of drug-kinase pairs.

    \xhdr{Other Binding Affinity Datasets}
        \label{subsec:other_datasets}
        AffinDB \citep{block2006} was a major binding affinity dataset when it was proposed in 2006. However, the website and affinity data have not been updated for some time. As of 2019, there are 748 affinity values covering 474 PDB complexes, which is significantly fewer than PDBbind and BindingDB.
        CSAR (Community Structure-Activity Resource) \citep{smith2011, dunbar2011, dammganamet2013, dunbar2013, smith2016, carlson2016} was an experimental dataset of crystal structures and binding affinities for diverse protein-ligand complexes. The dataset was intended for a community-wide exercise conducted by a group at the University of Michigan from 2010 to 2014, aiming to use unpublished data from in-house projects to evaluate existing protein-ligand binding affinity prediction methods. CSAR has inspired some of the most meaningful discussions on the methods and evaluations of binding affinity~\citep{novikov2011, koes2013, carlson2016_lessons}, which remain relevant today. %
        Aside from these unmaintained datasets, there are many datasets that are relevant to binding affinity. 
        For instance, \citet{tang2014} cite studies of  bioactivity profiling of small-molecule protein kinase inhibitor by \citet{davis2011}, \citet{metz2011}, and  \citet{anastassiadis2011}, each of which provides a set of measurements of binding affinity between kinase and small compounds.
        \citet{krivk2018} combined multiple small protein-ligand complex datasets, such as CHEN11 \citep{chen2011}, COACH420 \citep{roy2012_cofactor, yang2013_site}, and HOLO4K \citep{schmidtke2010}, for training and evaluation of binding site prediction. {
        CrossDocked2020 \citep{francoeur2020three}, which contains 22.5 million ligand poses docked into diverse but structurally related binding pockets from the Protein Data Bank, is also widely employed as a benchmark dataset for evaluating protein–ligand binding affinity prediction methods.
        }
        {
        As existing datasets and models fail to account for the dynamic features of protein-ligand interactions, the recently released PLAS-20K (2024) addresses this gap. As an extension of PLAS-5K (2022), PLAS-20K includes 97,500 independent simulations across 19,500 different protein-ligand complexes.
        Nevertheless, PLAS-20K remains one of the few publicly available molecular dynamics resources, and its scale is still modest compared to static datasets such as PDBbind. Most other MD-based datasets are either proprietary or highly heterogeneous, limiting their utility as community benchmarks. This scarcity of large, standardized dynamic datasets represents a key bottleneck, as models trained only on static snapshots often fail to capture critical factors such as protein flexibility, induced fit, and entropic effects.

        While PDBbind has been the de facto benchmark dataset for binding affinity prediction, it is well recognized that it suffers from significant issues of data leakage and redundancy, which limit the generalizability of models trained on it. This concern becomes particularly evident when evaluating models on proprietary or out-of-distribution industrial datasets, where performance often drops substantially. Recent efforts have emphasized the need for more robust benchmarks that explicitly address data leakage and bias. For instance, initiatives such as Plinder and discussions in the community \citep{omsf2023benchmarks} have highlighted directions for curating cleaner, more diverse datasets with stricter train–test splits to improve real-world applicability. Moreover, some traditional resources like Binding MOAD are no longer actively maintained \citep{bindingMOADdefunct}, further underscoring the importance of developing next-generation benchmark datasets. Future progress in binding affinity prediction will thus depend not only on improved modeling techniques but also on the availability of high-quality datasets designed to avoid redundancy and leakage.
        }

\section{Methodology 
}
\label{sec:4}

    Next, we discuss the various protein-ligand binding affinity prediction methodologies developed from the early 2000s to the present day.
    We divided all the methods into two categories based on model complexity and input data.
    (1) Conventional methods, which are essentially a set of energy equations derived from assumptions and understanding of the binding process, combining weighted physio-chemical terms in an additive manner into a single estimate for affinity.
    (2) Machine learning models (e.g., random forest, support vector machine, neural network) trained with human-engineered descriptors extracted from the protein-ligand complexes; and representation-learning methods, which extract features directly from raw data of protein-ligand complexes (e.g., SMILES strings, voxels, graphs) using the learning capacity of deep neural networks.
    As the study of binding affinity advances, the community has shifted from methods that require extensive domain knowledge and assumptions to those that can exploit the increasing number of available protein-ligand complex structures.

    \subsection{Conventional Methods}

    Since the formalization of the concept of protein-ligand docking,
    researchers have been trying to predict the binding affinity or energy based on human understanding of physics and chemistry. 
    These methods were often called scoring functions, rather than predictors or models, because they almost always tool the forms of additive functions, combining various engineered physio-chemical terms. 
    Traditionally, these conventional scoring functions have been roughly divided into the following three categories \citep{cheng2009, ballester2010, ashtawy2012}: Physics-based methods, empirical methods, and knowledge-based potential methods. 
    Among these, physics-based methods represent the most direct attempt to calculate binding affinity from first principles, and thus form a natural starting point for our discussion.
    
    \subsubsection{Physics-based methods}

   Traditional calculations of protein--ligand binding affinity that are grounded in statistical mechanics can be divided into two main categories. Both ultimately connect to the standard definition of the binding free energy,
\begin{equation}
  \Delta G_b^\circ = -k_{\mathrm{B}}T \ln\!\big(C^\circ K_b\big),
  \label{eq:std}
\end{equation}
where $K_b$ is the equilibrium association constant and $C^\circ$ is the 1~M standard concentration.

The first category is commonly referred to as free energy surface (FES) or potential of mean force (PMF) approaches. These methods compute a reduced-dimensional free energy profile along selected reaction coordinates that capture the essential physics of the binding process. The resulting profile reveals minimum free-energy pathways and critical points such as intermediates and transition states, from which both equilibrium and kinetic information can be extracted. In the simplest radial case, the binding constant can be obtained from the PMF as
\begin{equation}
  K_b = \int_{\text{site}} 4\pi r^2 \, e^{-\beta[\,w(r)-w(r^\ast)\,]} \, dr ,
  \label{eq:pmf}
\end{equation}
where $w(r)$ is the PMF and $r^\ast$ is a reference position in bulk solvent. 

The second category is alchemical free energy (AFE) methods \citep{ngo2024alchemical}. Here, the focus is not on the real physical pathway but on exploiting the fact that free energy is a state function. The ligand is gradually “alchemically” decoupled from its surroundings, both in the binding site and in bulk solvent, and the difference provides the binding free energy. A general expression can be written as
\begin{equation}
  \Delta G_b^\circ =
    \big[\Delta G^{\text{site}}_{\text{int}} - \Delta G^{\text{bulk}}_{\text{int}}\big]
    + \Delta G_{\text{restraints}},
  \label{eq:alch}
\end{equation}
where the first term accounts for the difference in interaction free energies between site and bulk, and $\Delta G_{\text{restraints}}$ includes standard-state, translational, rotational, and conformational corrections. Alchemical methods are often more computationally efficient and robust for estimating $\Delta G_b^\circ$, but they sacrifice mechanistic insight into the actual binding pathway.

\textbf{Relative binding free energy (RBFE) methods.} While the alchemical approach
refers to absolute binding free energy (ABFE) calculations, in practice the \emph{relative} binding free 
energy method has become far more widely adopted, particularly in industrial drug discovery settings. Rather 
than computing the binding affinity of a single ligand from scratch, RBFE calculates the free energy 
\emph{difference} between two structurally similar ligands binding to the same target:
\begin{equation}
\Delta\Delta G = \Delta G_{\text{b}}^{\text{ligand B}} - \Delta G_{\text{b}}^{\text{ligand A}}.
\end{equation}
This is achieved by performing alchemical transformations that mutate ligand A into ligand B, both in the 
binding site and in solvent. The relative formulation dramatically reduces the sampling burden: because 
structurally similar ligands often occupy overlapping conformational spaces and induce similar protein 
reorganization, many systematic errors and convergence challenges cancel out in the difference. This makes 
RBFE substantially more computationally tractable than ABFE for comparing congeneric series of 
compounds~\citep{wang2015accurate, cournia2017relative}.

As a result, RBFE remains the gold standard for prospective ligand optimization in pharmaceutical settings, 
with demonstrated success in predicting relative potencies with chemical accuracy (± 1 kcal/mol) 
\citep{behera2025quantification}.

In practice, however, both FES- and AFE-based approaches are limited by the inherently low sampling
efficiency of molecular dynamics simulations \citep{york2023modern}. As binding often involves slow conformational
changes and rare transitions, straightforward Boltzmann sampling cannot ensure sufficient exploration of
the relevant configurational space. 

To overcome this challenge, these methods are commonly combined with
enhanced sampling techniques, such as replica exchange \citep{sugita1999replica}, umbrella sampling \citep{torrie1977nonphysical}, or metadynamics \citep{laio2002escaping}, which improve convergence by facilitating transitions across free energy barriers and ensuring
adequate sampling of binding-relevant degrees of freedom.

However, exhaustive sampling of all relevant conformational states is often prohibitively 
expensive for realistic biomolecular systems. This limitation has motivated the development 
of so-called \textit{endpoint methods}, which bypass the explicit sampling of the binding 
pathway by considering only the end states: the bound complex and the separated receptor 
and ligand. Among these, the MM/PBSA and MM/GBSA \citep{kollman2000calculating} approaches have become particularly 
popular, owing to their balance between computational cost and predictive accuracy. In these 
methods, the binding free energy is expressed as

\begin{equation}
\Delta G_{\mathrm{bind}} = \Delta E_{\mathrm{MM}} + \Delta G_{\mathrm{solvation}} - T \Delta S ,
\end{equation}
where $\Delta E_{\mathrm{MM}}$ contains bonded, electrostatic, and van der Waals terms, 
$\Delta G_{\mathrm{solvation}}$ is decomposed into polar and non-polar contributions, and 
$T \Delta S$ represents the entropic term.

The distinction between MM/PBSA and MM/GBSA lies in the treatment of the polar solvation 
energy. In MM/PBSA, it is obtained by numerically solving the Poisson--Boltzmann (PB) 
equation, whereas MM/GBSA employs the Generalized Born (GB) approximation:
\begin{equation}
\Delta G_{\mathrm{solvation}} = G_{\mathrm{pol}}^{\mathrm{PB/GB}} + G_{\mathrm{np}}^{\mathrm{SASA}} .
\end{equation}

Here, the polar solvation free energy, $G_{\mathrm{pol}}^{PB/GB}$, is obtained either by 
numerically solving the Poisson--Boltzmann (PB) equation or by employing the Generalized Born (GB) 
approximation. In the PB formulation, the electrostatic potential $\phi(\mathbf{r})$ is obtained from
\begin{equation}
-\nabla \cdot \left[ \epsilon(\mathbf{r}) \nabla \phi(\mathbf{r}) \right] + \kappa^2(\mathbf{r}) \sinh \!\left( \phi(\mathbf{r}) \right) 
= 4 \pi \rho(\mathbf{r}),
\end{equation}
and the corresponding solvation energy is calculated as
\begin{equation}
G_{\mathrm{pol}}^{PB} = \frac{1}{2} \sum_i q_i \, \phi(\mathbf{r}_i).
\end{equation}

In the GB approximation, the polar solvation energy is estimated analytically as
\begin{equation}
G_{\mathrm{pol}}^{GB} = -\frac{1}{2}\left(1 - \frac{1}{\epsilon}\right) 
\sum_{i=1}^N \sum_{j=1}^N \frac{q_i q_j}{f_{ij}},
\end{equation}
with
\begin{equation}
f_{ij} = \sqrt{ r_{ij}^2 + \alpha_i \alpha_j \exp\!\left(-\frac{r_{ij}^2}{4 \alpha_i \alpha_j}\right) },
\end{equation}
where $r_{ij}$ is the interatomic distance, and $\alpha_i$ are the effective Born radii.

The principal advantages 
of these endpoint approaches are their efficiency and modularity: they require only molecular 
mechanics snapshots, can be combined with standard MD simulations, and allow energy 
decomposition analysis. Nevertheless, their accuracy is strongly system-dependent, with 
limitations arising from approximate entropy estimates, neglect of conformational 
reorganization and water thermodynamics, and sensitivity to continuum solvent parameters. 
Thus, while MM/PBSA and MM/GBSA represent an attractive compromise between speed 
and accuracy, their predictive power remains limited compared to fully converged 
alchemical free energy methods. While physics-based methods strive to compute binding free energies directly from first principles, their high computational cost and dependence on approximations have led to the development of empirical methods, which instead introduce experimentally motivated terms and statistical regression to improve efficiency and practical accuracy.

    \subsubsection{Empirical methods.}

        Empirical methods are the most commonly used score-based models, found in packages such as Rosetta and Autodock \citep{rohl2004protein_rosetta, trott2010autodock}. In addition to the binding energy terms from the previous physics-based methods, they include contributions from empirical chemistry factors, such as hydrophobicity, metal-ligand interactions, entropy effects from steric hindrance, and ligand motifs \citep{rohl2004protein_rosetta, o2015combined-rosetta, park2016simultaneous-rosetta, alford2017rosetta, halgren2004glide, friesner2006extra-glide, quiroga2016vinardo}.

        These methods employ statistical learning algorithms, such as  multivariate linear regression (MLR) or partial least squares regression (PLS), to parameterize the individual energy terms with weights for estimating the binding affinity. %
        Generally, empirical methods are more flexible and adaptive, allowing users to add customized terms to the scoring function. They adopt a generalized functional form:
        \begin{equation}
            \mathrm{Score} = E_\mathrm{bind} + E_\mathrm{emp}
        \end{equation}
        where $E_{bind}$ is binding affinity from the physics model and $E_{emp}$ represents empirical terms that account for interactions and contributions that pure physics fails to capture due to the limitations of force field parameters.

        \paragraph{Hydrophobic Interaction Terms}
        Hydrophobic interactions constitute one of the most critical empirical components. AutoDock Vina implements hydrophobic terms using a distance-dependent step function \citep{trott2010autodock}:
        \begin{equation}
        E_\mathrm{hydrophobic} = w_\mathrm{hydrophobic} \sum_{i,j} f_\mathrm{hydrophobic}(r_{ij}, t_i, t_j)
        \end{equation}

        \paragraph{Metal-Ligand Interaction Terms}
        Metal coordination represents a specialized empirical term crucial for metalloproteins. The general form incorporates both distance and angular constraints \citep{santos2014autodock4zn}:
        \begin{equation}
        E_\mathrm{metal} = \sum_{M,L} w_\mathrm{ML} \cdot g(r_\mathrm{ML}, r_\mathrm{opt}, \sigma) \cdot h(\theta, \theta_\mathrm{ideal})
        \end{equation}
        where the distance function follows a Gaussian form:
        \begin{equation}
        g(r_\mathrm{ML}, r_\mathrm{opt}, \sigma) = \exp\left(-\frac{(r_\mathrm{ML} - r_\mathrm{opt})^2}{2\sigma^2}\right)
        \end{equation}
        and the angular function penalizes deviations from ideal coordination geometry:
        \begin{equation}
        h(\theta, \theta_\mathrm{ideal}) = \exp\left(-\alpha(\theta - \theta_\mathrm{ideal})^2\right)
        \end{equation}

        \paragraph{Entropy Effect Terms}
        Empirical entropy terms account for the loss of conformational and rotational freedom upon binding. The rotational entropy loss is approximated using the rigid rotor model \citep{gilson1997statistical}:
        \begin{equation}
        \Delta S_\mathrm{rot} = -R \ln\left(\frac{8\pi^2 I_A I_B I_C}{\sigma_\mathrm{rot} h^3}\right)
        \end{equation}
        where $I_A$,  $I_B$, $I_C$ are the principal moments of inertia, $\sigma_{rot}$ is the rotational symmetry number, and $h$ is Planck's constant.
        
        For conformational entropy, AutoDock Vina employs a simplified model based on the number of rotatable bonds \citep{trott2010autodock}:
        \begin{equation}
        E_\mathrm{entropy} = w_\mathrm{tors} \cdot \frac{N_\mathrm{tors}}{1 + w_\mathrm{tors} \cdot N_\mathrm{tors}}
        \end{equation}
        where $N_{tors}$ is the number of rotatable bonds and $w_{tors}$ represents the entropic penalty per rotatable bond.

        The latter additional terms improve the binding affinity prediction based on pure physics models and can be tuned to achieve a better fit for particular interests, such as pharmaceutical importance \citep{wang2004extensive_comparison, friesner2006extra-glide}.
        However, empirical methods require substantial empirical knowledge to set up appropriate scoring functions and extensive training datasets to optimize the weight of each term. The challenge lies in balancing model complexity with generalizability, as overfitting to training data can compromise performance on novel chemical scaffolds or binding sites \citep{li2014comparative}. While empirical methods enhance physics-based scoring by incorporating experimentally motivated terms and statistical regression, they still require substantial expert knowledge to define and parameterize interaction terms. To reduce this dependence on manual feature design, researchers developed knowledge-based potential methods that instead derive interaction preferences directly from large databases of protein–ligand complexes.

    \subsubsection{Knowledge-based potential methods.} 
        Knowledge-based potential methods rely on learning from a database of protein-ligand complexes to determine the potential between atom pairs and predict binding affinity. The main assumption behind these methods is that if an atom pair, one from the ligand and one from the protein, appears with a higher frequency than the reference distribution, it might indicate an energetically favorable interaction between the given pair. Thus, the distance-dependent potential of an atom pair can be obtained through inverse Boltzmann analysis based on the measured occurrence frequency of this pair across the entire knowledge base. These methods are sometimes referred to as \textit{knowledge-based methods}~\citep{gohlke2000predicting} or \textit{potential of mean force methods}~\citep{su2009quantitative}.
        The general functional form of knowledge-based potential methods is represented by:
        \begin{equation}
            \text{Score} = \sum_{i \in \text{ligand}} \sum_{j \in \text{protein}} - k_B T \ln{\big( \frac{\rho_{ij} (r)}{\rho^*_{ij}}  \big)}
        \end{equation}
        where $k_B$ is the Boltzmann constant, $T$ is the absolute temperature, and $r$ is the distance between pairs of atoms.
        Knowledge-based methods are more general, implicitly incorporating effects not fully understood from the structural data. They can also incorporate some physics-based or empirical terms to enhance performance.
        
       Knowledge-based potential methods offer several distinct \textit{advantages}:
       
(1) Computational efficiency: These methods are computationally lightweight and fast, making them suitable for high-throughput screening applications where rapid evaluation of large compound libraries is essential.

However, these methods face significant \textit{limitations}:

(1) Structural bias over energetic relevance: The statistical distributions of atom-pair distances are primarily reflective of geometrically favorable binding poses rather than true thermodynamic binding affinity. This theoretical limitation suggests that knowledge-based potentials may not capture the energetic determinants of binding strength. Paradoxically, empirical evaluations demonstrate that these methods often perform comparably to more sophisticated approaches, indicating that structural complementarity may be a reasonable proxy for binding affinity in many cases.

(2) Limited physical interpretability: While knowledge-based potentials represent a significant advancement by introducing data-driven elements to complement purely physics-based or empirical approaches, they remain fundamentally phenomenological. The statistical potentials lack clear physical meaning and cannot easily be related to specific intermolecular forces or thermodynamic properties.
Recognition of these complementary strengths and limitations has motivated the development of hybrid scoring functions that strategically combine knowledge-based potentials with physics-based calculations and empirical observations, aiming to harness the computational efficiency of statistical methods while maintaining physical rigor and predictive accuracy.

    \subsubsection{Hybrid methods} %
    Efforts have been made to bring different categories of scoring functions together to improve performance, blurring the boundaries between these categories. Some physics-based methods use weight parameters derived from regressions to increase performance. Furthermore, some knowledge-based potential methods add solvation and entropy terms \citep{liu2015_sf_clf}.
    
    Multiple reviews and comparative studies have been conducted to assess the conventional scoring functions. 
    A comparative study of 16 popular conventional scoring functions \citet{cheng2009} indicates that no single one consistently outperforms the others in scoring, ranking, and docking. 
    The Pearson correlation between predicted scores from the most commonly used scoring functions (e.g.,~Glide~\citep{halgren2004glide}, AutoDock, Dock) and the experimentally-determined binding affinity ranges from 0.4 to 0.6. This evidence provides some justification for the use of docking, but it is not sufficient to be applied in virtual screening and drug discovery with high confidence. 
    Despite the subpar performance in the prediction of binding affinity, studies have found that binding software with conventional scoring functions can accurately predict the correct binding pose and conformation (60-80\%)~\citep{cheng2009, plewczynski2011, liyan2014_2}. This precision is typically not achieved by descriptor-based scoring functions~\citep{xie2014}.
    
    Conventional scoring functions were proposed in the early stages of research on structure-based virtual screening and are widely used in almost all commercial and academic docking software.
    These scoring functions are still useful and relevant to research on protein-ligand binding today. Most of them are based, to some extent, on the theory of physics and chemistry, which makes them not only reliable and explainable to some degree but also capable of improvement as the theory advances. \citet{cavasotto2020high} shows that incorporating a PM7 semiempirical quantum mechanical method  as a scoring function significantly improves the number of compounds correctly screened using molecular docking. 
    Another advantage of conventional scoring function is that---despite the use of statistical machine learning algorithms---their simple function form and low learning capacity allow them to perform reasonably well without training on too much data. 
    This characteristic was extremely desirable before and during the 2000s, when there were not enough complex structures available for the training of complex learning algorithms and models.
    
    Yet despite these advantages and some recent progress, conventional scoring functions are losing momentum in both research and industry.
    The disadvantages of conventional scoring functions are mostly rooted in their additive functional forms \citep{li2015}.
    First, the energy/potential terms in the additive functions are assumed to be independent with each other, which is often not the case, especially for scoring functions based on energy terms \citep{khamis2015}.
    Secondly, these functions are not expressive enough for non-linearity, indicating their incapability of complex curve-fitting and conditional branching.
    As such, these scoring functions cannot take advantage of the growing number of
available higher-quality protein-ligand complex structures data \citep{ballester2014}.
    Lastly, even from a theoretical point of view, upon which the conventional methods were usually based, these scoring functions ignored many important aspects of binding, such as the implicit treatment of solvent and protein flexibility \citep{ballester2010}, which became sources of error that are difficult to address.
    All these drawbacks are related to two factors that are deeply rooted in all conventional scoring functions: (1) the underlying theory of docking is incomplete, and the scoring functions, as a way to approximate the binding affinity, are flawed \citep{pantsar2018}; (2) the additive functional form neither accurately reflects the physio-chemical process of binding nor is suitable for learning from the increasing number of available protein-ligand complex structures.

    Conventional scoring functions provided the first generation of computational predictors, balancing efficiency and interpretability. Yet, as structural databases expanded and computational power increased, their rigid formulations proved inadequate for capturing the complexity of protein–ligand interactions, thereby catalyzing the rise of machine learning–based methodologies.

\subsection{Machine Learning-based Methods}

The prohibitive cost and scalability limits of conventional physics-based approaches (e.g., OpenEye) have motivated increasing reliance on machine learning for binding affinity prediction. Within this paradigm, two complementary directions have emerged: interaction-free models, which are primarily representation-driven and infer affinities from learned embeddings of proteins and ligands without explicit structural interactions, and interaction-based models, which are more physics-informed and ground their predictions in the spatial and chemical features of binding pockets. {In addition, during prediction, a variety of representations are employed to capture the structural and chemical characteristics of proteins and molecules. Sequence-based approaches model proteins as amino acid sequences and molecules as SMILES strings, emphasizing their primary structural information. Graph representations encode proteins by treating atoms as nodes with edges formed via k-nearest neighbors from 3D coordinates, while molecules are represented with atoms as nodes and chemical bonds as edges, capturing topological relationships. Voxel-based methods discretize protein structures into 3D grids that encode the spatial arrangement of binding sites for convolutional modeling. Finally, point cloud representations describe proteins as sets of atomic coordinates in 3D space, facilitating direct geometric reasoning over spatial distributions.} This dichotomy reflects a broader philosophical divide between abstract representation learning and explicit interaction modeling, framing much of the recent methodological innovation in the field.

\subsubsection{Interaction-free approach}

Interaction-free models infer binding affinity from data without focusing on direct physical interactions. Specifically, these ML-based models consist of two separate parts, each aiming to learn representations from molecule and protein data, including SMILES, protein sequences and graphs. The interactions between proteins and ligands are implicitly captured in the latent spaces of their embeddings, which are formed through a neural network that processes their concatenated representations.

Early models in this field used 1D sequence-level representations, such as SMILES for molecules and protein sequences for proteins \citep{ozturk2018deepdta, abbasi2020deepcda, wang2021deepdtaf, yang2021ml, zeng2021deep, zhao2022attentiondta, yuan2022fusiondta}. These early models were innovative but had limitations because they did not include important information about the 3D structures of molecules and proteins. This lack of 3D information often led to less accurate predictions of binding affinities.

With the development of graph neural networks (GNNs), researchers began using more advanced representations. 
They started by representing molecules as {2D molecular graphs, where atoms are nodes and bonds are edges, thereby capturing the topological relationships between atoms. Proteins, however, are still typically represented as linear amino acid sequences  \citep{nguyen2021graphdta, yang2022mgraphdta}}. This change allowed for more accurate modeling of molecular structures and improved the models' predictions.

As GNNs improved, researchers began representing complex protein 3D structures as graphs too. This approach, where both proteins and molecules are represented as graphs, has greatly enhanced the models' ability to capture the details of molecular interactions \citep{somnath2021holoprot, jiang2022sequence, wang2023pronet}. By including spatial information, these models enhances representation fidelity by explicitly modeling 3D topological and spatial constraints of protein-ligand complexes, leading to better predictions of binding affinity.

Additionally, the growing interest in multimodal models has led to the development of multimodal models that combine both 1D sequences and 3D graphs to represent proteins \citep{wu2023integration, zhang2023multimodal, wu2024attentionmgt, liu2025bidirectional}. These models take advantage of the strengths of different data types, combining detailed sequence information from 1D sequences with spatial and relational information from 3D graphs. This combination is especially useful for capturing complex interactions that might be missed by using only one type of data. 
{Notably, large-scale pretrained protein language models have been increasingly adopted \citep{wu2023integration, wu2024attentionmgt, liu2025bidirectional}. These methods rely on pretrained protein language models such as ESM \citep{lin2023evolutionary_esm2} to derive embeddings from raw sequences. By leveraging large-scale pretraining on millions of protein sequences, these embeddings capture generalizable biochemical and evolutionary features, which can then be integrated with structural representations. Such pretraining strategies have been shown to improve generalization to unseen proteins, underscoring the scalability and adaptability of interaction-free approaches within multimodal settings.}

\begin{table}[ht]
\centering
\begin{tabular}{llll}
\toprule
\textbf{Models} & \textbf{Protein Repr} & \textbf{Molecule Repr} & \textbf{Dataset  Used} \\
\midrule
\citep{ozturk2018deepdta} & Sequence & Sequence & Davis, KIBA  \\
\hline
\citep{abbasi2020deepcda} & Sequence & Sequence & Davis, KIBA, BindingDB \\
\hline
\citep{wang2021deepdtaf} & Sequence & Sequence & PDBBind\\
\hline
\citep{yang2021ml} & Sequence & Sequence & Davis, KIBA, Metz \\
\hline
\citep{zeng2021deep} & Sequence & Sequence & Davis, KIBA \\
\hline
\citep{zhao2022attentiondta} & Sequence & Sequence & Davis, KIBA, Metz  \\
\hline
\citep{yuan2022fusiondta} & Sequence & Sequence & Davis, KIBA  \\
\hline
\citep{nguyen2021graphdta} & Sequence & Graph & Davis, KIBA  \\
\hline
\citep{yang2022mgraphdta} & Sequence & Graph & \makecell[l]{Davis, KIBA, Metz} \\
\hline
\citep{somnath2021holoprot} & Graph & Graph & PDBBind  \\
\hline
\citep{jiang2022sequence} & Graph & Graph & Davis, KIBA  \\
\hline
\citep{wang2023pronet} & Graph & Graph & PDBBind  \\
\hline

\citep{wu2023integration} & \makecell[l]{Multi (Graph + Seq)} & Graph & PDBBind  \\
\hline
\citep{zhang2023multimodal} &\makecell[l]{Multi (Graph + Seq)} & \makecell[l]{Multi (Graph + Seq)} & Davis, KIBA  \\
\hline
\citep{wu2024attentionmgt} & \makecell[l]{Multi (Graph + Seq)} & Graph & Davis, KIBA  \\
\hline
\citep{liu2025bidirectional} & \makecell[l]{Multi (Graph + Seq)} & Graph & PDBBind  \\
\bottomrule
\end{tabular}
\caption{Comparison of Models for Interaction-free Binding Prediction}
\label{table:interaction_free_binding_models}
\end{table}

However, even with these advancements, interaction-free models face challenges. The inputs to these models often lack detailed interaction information, making it difficult to accurately predict protein-ligand binding affinity. Moreover, GNNs often struggle to capture essential long-range interaction information between proteins and ligands, which is crucial for predicting binding affinity accurately.

\subsubsection{Interaction-based approach}
Interaction-based models make predictions based on the 3D structures of complexes and the physical interactions between proteins and ligands. These models employ only the atoms surrounding the interaction/binding pocket to build graphs for prediction.

Interaction-based methods are primarily dominated by 3D voxel grids and graphs, relying on 3D CNNs and GNNs, respectively.
3D voxel grid-based methods \citep{jimenez2018k, li2019deepatom, hassan2020rosenet, jones2021improved} use a 3D voxel grid to represent the 3D structures of protein-ligand complexes as input features for CNNs. However, the 3D voxel grid representation has several limitations. First, it has high memory consumption and computational cost due to the sparsity of the voxels, as the protein structure occupies only a small part of the entire grid. Additionally, the sensitivity of 3D grids to rotation negatively impacts prediction results. 
In contrast, graph-based methods \citep{jiang2021interactiongraphnet, li2021signdta, moon2022pignet, yang2023gigndta, yu2023computing} are mostly rotation-invariant, making their graph representations more robust than grid representations.
Still, interaction-based models are limited to known protein-ligand complex structures, making them less useful compared to interaction-free methods when faced with unknown protein-ligand complexes.

{To address these challenges, recent research has begun to develop models capable of jointly predicting high-quality protein–ligand complex structures and their binding affinities \citep{lu2022tankbind, tan2024gaabind}. Such approaches aim to bridge the gap between interaction-free and interaction-based paradigms, offering the potential to generate reliable predictions even in the absence of experimentally resolved complexes. Although there are works combining complex prediction and binding affinity prediction, {most recent structure prediction models \citep{alphafold_jumper2021highly, krishna2024generalize_rfaa, abramson2024accurate_af3, liu2024technical_helixFold3}
while able to predict complex structures, are not designed to predict binding affinities.} In contrast, the most recent Boltz2 \citep{passaro2025boltz} framework supports both binding affinity prediction and complex structure prediction, highlighting a step forward in unifying these tasks. 
}

{Beyond the methodological advances, another emerging direction is the integration of dynamics and flexibility into complex representations. Most existing approaches treat protein–ligand interactions as static snapshots, neglecting the conformational variability that can critically influence binding affinity. Recent works have begun to explore incorporating molecular dynamics simulations \citep{min2024static, passaro2025boltz} or {diffusion-based generative frameworks \citep{guan20233d, chai2024chai, wohlwend2025boltz1, bytedance2025protenix, lin2025diffbp, passaro2025boltz}} to model the ensemble nature of protein–ligand complexes. By explicitly accounting for conformational changes and induced fit effects, these approaches aim to better approximate real binding processes and improve generalization to diverse biological systems. Such dynamic-aware methods represent a promising avenue for bridging the gap between computational predictions and experimentally observed binding behaviors.}

\begin{table}[ht]
\centering
\begin{tabular}{lll}
\toprule
\textbf{Models} & \textbf{Repr} & \textbf{Dataset Used} \\
\midrule
\citep{jimenez2018k} & Voxel & PDBbind  \\
\hline
\citep{li2019deepatom} & Voxel & PDBBind\\
\hline
\citep{hassan2020rosenet} & Voxel & PDBBind \\
\hline
\citep{jones2021improved} & Voxel & PDBBind  \\
\hline
\citep{jiang2021interactiongraphnet} & Graph  & PDBBind \\
\hline
\citep{li2021signdta} & Graph & PDBBind,  CSAR \\
\hline
\citep{moon2022pignet} & Graph & PDBBind, CASF, CSAR \\
\hline
\citep{lu2022tankbind} & Graph & PDBBind \\
\hline
\citep{yang2023gigndta} & Graph & PDBBind, CSAR\\
\hline
\citep{yu2023computing} & Graph & BindingDB \\
\hline
\citep{guan20233d} & Graph & PDBBind, {CrossDocked2020} \\
\hline
\citep{tan2024gaabind} & Graph & PDBBind, CASF \\
\hline
\citep{passaro2025boltz} & {Point cloud} & {PubChem, ChEMBL, BindingDB} \\
\bottomrule
\end{tabular}
\caption{Comparison of Models for Interaction-based Binding Prediction}
\label{table:interaction_based_binding_models}
\end{table}

Advances in both interaction-free and interaction-based machine learning approaches signal a decisive shift beyond the limitations of conventional scoring functions, with predictive improvements emerging alongside broader transformations in drug discovery, such as the FDA’s recent phase-out of mandatory animal testing. Within this evolving landscape, AI-driven in silico frameworks—most notably AI Virtual Cells (AIVCs)—are poised to redefine binding affinity prediction by situating molecular interactions within dynamic, multi-omic, and cell-type–specific contexts. Progress in affinity prediction will directly enhance the fidelity of AIVCs, while advances in AIVCs will reciprocally provide richer system-level environments to refine and validate predictive models, creating a mutually reinforcing cycle that could ultimately enable simulations of temporal dynamics, system-level specificity, and more personalized therapeutic outcomes. Yet, critical gaps remain, particularly in addressing conformational flexibility, dataset bias, and the integration of multi-omics information—challenges we explore further in Section~\ref{section:discussion}.

\section{Evaluation}
\label{sec:3}
    
    Different from most numeric regression problems, the evaluation of binding affinity prediction is much more complex than simple error assessment, which has been pointed out repeatedly in multiple comparative studies and benchmarking papers \citep{cheng2009, ashtawy2012, liyan2014_2, khamis2015_comparative, liu2017, su2018}.
    Evaluation by ranking with decoy ligands and binding poses is necessary to simulate the process of virtual screening, ensuring the model's practical usefulness.
    \citet{cheng2009} evaluated scoring functions or binding affinity predictors from three perspectives: \textit{scoring power} (binding affinity prediction), \textit{docking power} (binding pose prediction), and \textit{ranking power} (ligands relative ranking prediction). 
    In a more recent study by the same group, \citet{liyan2014_2} enriched the evaluation by adding \textit{screening power} (discrimination of true binders from decoys) to the set.    
    
    \subsection{Scoring Power}
        
        Scoring power~\citep{su2020tapping} refers to the ability to produce binding scores that are correlated with the experimentally-measured affinities, preferably in a linear fashion. 
        For conventional scoring functions, the scoring power could be measured with Pearson’s correlation coefficient ($R$) or standard deviation ($\mathrm{SD}$):
        \begin{align}
            R = \frac{\sum (x_i - \mean{x}) (y_i - \mean{y})}{\sqrt{\sum {(x_i - \mean{x})}^2} \sqrt{\sum {(y_i - \mean{y})}^2}}, \quad \mathrm{SD} = \sqrt{\frac{\sum {[y_i - (a x_i + b)]}^2}{N - 1}}
        \end{align}
        where $x_i$ is the prediction for the $i$th complex in the evaluation set, $y_i$ is the experimental binding affinity, $a$ and $b$ are linear regression factors between the predicted scores and binding affinity, and $N$ is the number of samples in the evaluation set.

        However, machine learning methods can often estimate the binding affinity directly instead of producing a score.
        In these cases, common regression metrics such as Mean Absolute Error (MAE), Mean Squared Error (MSE) or Root Mean Squared Error (RMSE) can also be used.
        Still, Pearson's correlation coefficient ($R$) seems be the most commonly used one for its simplicity and invariance to scaling and unit.
        There exists other statistical indicators that reflect the linear correlation between the predictions and targets.
        However, these indicators often correlate with Pearson's correlation coefficient and do not provide extra information.
        
        Scoring power is the most essential ability of binding affinity models, which means that the evaluation of scoring power is often adequate for studies that focus solely on affinity prediction. 
        However, in most cases, we are also curious about the models' performance in different settings, such as docking pose prediction and virtual screening.
        It is more than common for a model to make excellent predictions for binding affinity, but be incapable of differentiating between the true ligand from decoys, ultimately rendering the model useless in drug discovery \citep{liyan2014_2}. 
        As such, other evaluations for docking, ranking, and screening are required to assess the models in a more comprehensive way.

    \subsection{Docking Power}
    
        Docking power refers to the ability of a model to differentiate between the true binding pose from the decoy poses of a given protein-ligand pair.
        Ideally, the complex with the native binding pose should have the highest score or predicted affinity compared to the decoys. 
        To implement such evaluation, complexes with decoy binding poses are usually generated with a molecular docking program or through molecular dynamics simulation. 
        The later versions of CASF \citep{liyan2014_2} used multiple docking programs (e.g.,~AutoDock \citep{huey2012using}, LigandFit \citep{venkatachalam2003ligandfit}, GOLD \citep{verdonk2003improved}, Surflex \citep{jain2003surflex}, FlexX \citep{schellhammer2004flexx}) for decoy pose generation to minimize the bias in conformation sampling. 
        The native/true pose is explicitly added into the decoy set so that there is at least one correct binding pose in CASF \citep{cheng2009}.
        After the scores/affinities for the decoy poses are generated, poses with the highest score/affinity are compared against the true pose using RSMD, which is calculated with:
        \begin{equation}
            \text{RMSD} = \sqrt{\frac{\sum [{(x_i -x'_i)}^2 + {(y_i -y'_i)}^2 + {(z_i -z'_i)}^2]}{N}}
        \end{equation}
        where $(x_i, y_i, z_i)$ and $(x'_i, y'_i, z'_i)$ are the coordinates of the $i$th atom in the true and predicted decoy poses. Lower RMSD indicates that the predicted pose is closer to the native one.

        Using a static pose directly without a cutoff might cause misleading evaluation results: (1) due to the resolution and different structure measurement techniques, the native poses obtained experimentally might not be truly optimized; (2) we should consider the flexibility of proteins. 
        So, a model is considered to have successfully predicted the docking pose for a protein-ligand pair if the RMSD between the true and predicted poses is below a given cutoff.
        In CASF, the docking power of a model is evaluated by the success rate under different cutoffs ($1.0$, $2.0$, and $3.0$ \AA) for RMSD. 
        
        Still, using RMSD as a metric for distance deviation has drawbacks, such as sensitiveness to the atom ordering and unawareness of the flexibility of protein structures. 
        \citet{liyan2014_2} used RMSD\textsuperscript{PM}, which matches atom pairs between two poses using atom types instead of the atom IDs. 
        \citet{damm2006} proposed wRMSD, a weighted alternative to RMSD that takes into account the flexibility of proteins, which was then recommended by \citet{carlson2016_lessons} as a better alternative to naive RMSD.

    \subsection{Ranking Power}
    
        Ranking power refers to a model's ability to correctly rank the ligands based on their binding affinity to the given target protein. %
        A binding affinity scoring function/predictor may rank ligands with Spearman's rank correlation coefficient ($\rho$) or Kendall's rank correlation coefficient ($\tau$) as the evaluation metrics:
        \begin{align}
            \rho =& \frac{\sum (r_i - \mean{r}) (r'_i - \mean{r'}) }{\sqrt{\sum {(r_i - \mean{r})}^2 \sum {(r'_i - \mean{r'})}^2}} 
            = 1 - \frac{6 \sum {(r_i - r'_i)}^2}{n(n^2 - 1)} \\
            \tau =& \frac{2}{n(n-1)} \sum_{i<j} \text{sgn}(r_i - r_j) \text{sgn}(r'_i - r'_j)
        \end{align}
        where $r_i$ and $r'_i$ are the true and predicted rank of the $i$th member in the test set of size $n$.
        Besides these two commonly used statistical correlation coefficients, some other metrics are used to assess the rank power. 
        For instance, CASF adapts the predictive index (PI) proposed by \citet{pearlman2001} to measure rank, which places a higher weight on the complex pairs with significant differences in binding affinities:
        \begin{equation}
            \text{PI} = \frac{\sum_{i<j} [ (T_j - T_i) \cdot \text{sgn} (P_j - P_i) ] }{\sum_{i<j} | T_j - T_i | }
        \end{equation}
        where $T_i$ and $P_i$ are the target and predicted binding affinity for the $i$th test sample.
        Another example is CI (concordance index) \citep{gnen2005}, used to measure concordance of target and predicted ordering of binding affinities~\citep{ztrk2018, ztrk2019}:
        \begin{equation}
            \text{CI} = \frac{\sum_{T_i < T_j} [ 0.5 + 0.5 \cdot \text{sgn} (P_j - P_i) ] }{ \sum_{T_i < T_j} 1 }
        \end{equation}

    \subsection{Screening Power}
    
        Screening power~\citep{guedes2018empirical} refers to a model's ability to differentiate the true ligand binder from decoy molecules for a given protein target. 
        Screening power is computed by ranking all of the ligands in descending order of their score/affinity, and determining whether the true ligands are ranked highly, which is characterized with enhancement factor (EF):
        \begin{equation}
            \text{EF}_{\alpha} = \frac{\textit{NTB}_{\alpha}}{ \alpha \cdot \textit{NTB}_{\text{total}}}
        \end{equation}
        where $\alpha$ is an arbitrary percentage threshold, typically 1\%, 5\%, or 10\%; $\textit{NTB}_{\alpha}$ is the number of true binders found among the top $\alpha$ candidates; and $\textit{NTB}_{\text{total}}$ is the total number of true binders for a single target protein. %
        A major problem is the assumption that random ligands in the dataset do not bind with a target protein, which is often not the case. 
        For instance, \citet{su2018} report that 21 of the 57 target proteins ranked by CASF have more than five binding ligands in the dataset, according to known binding data in ChEMBL, which raises questions regarding CASF's evaluation for screening power. 
        Moreover, using random molecules as non-binders in a decoy set---although considered common practice in protein-ligand binding affinity and drug-target interaction modeling---can potentially hinder model learning and validation.

        Among the four evaluation aspects, scoring power seems to be the dominant factor, at least on the surface level. 
        Essentially, docking power, ranking power, and screening power are evaluated with the scores/affinities predicted from the model. 
        Thus, it is completely reasonable to assume that a binding affinity predictor with good scoring power might also perform well on the other three evaluations. 
        However, it may not be the case \citep{cheng2009}. 
        Ranking is a much harder task for conventional methods, and machine learning based predictors---despite their impressive results for scoring power---seem to perform poorly on docking, which renders the whole category of methods unconvincing.

\section{Discussion}\label{section:discussion}

We have provided an extensive review of binding affinity prediction methodologies, tracing their evolution from conventional approaches,
traditional machine-learning-based methods, to modern deep-learning-based methods. %
We also examine key challenges in binding affinity prediction, including dataset limitations such as measurement inconsistencies, chemical space biases, and incomplete experimental coverage, as well as the intrinsic complexity of protein-ligand interactions  that collectively impede accurate computational prediction.

Our review also leads us to identify five important areas for further research. 
1) \textit{Overcome limitations in existing datasets}. 
Current datasets are often biased towards protein-ligand pairs with favorable binding constants and correct poses, while complexes with low affinity or failed bindings are underrepresented. Future research could focus on creating more balanced datasets that better represent the full spectrum of binding affinities.
2) \textit{Integrate physics-based models with machine learning and deep learning}. There is potential for hybrid models that combine the theoretical rigor of physics-based methods with the predictive power of machine learning. Research in this area could explore how to better integrate these approaches to improve prediction accuracy.
3) \textit{Better handle protein flexibility}. Most existing models treat proteins as rigid entities, a simplification that overlooks important dynamics. Developing methods that account for protein flexibility, along with different ligand conformations during binding, could yield more accurate predictions.
4) \textit{Improve evaluation metrics}. Current evaluation metrics, such as Pearson correlation for scoring power and RMSD for docking power, have limitations. Research could explore alternative metrics that better capture the nuances of binding affinity predictions, particularly in real-world applications like drug discovery.
5) \textit{Multimodal data integration}. With the growing interest in multimodal models, future research could focus on integrating diverse data types (e.g., sequence data, 3D structures, interaction networks, molecular dynamics) to capture complex interactions more effectively.
Research in each of these areas can further advance the field of binding affinity prediction, potentially leading to more accurate and reliable tools for drug discovery and other applications.

Further, with the Food and Drug Administration~(FDA) recently phasing out animal testing~\citep{williams2024fda}, AI-driven tools—especially virtual cells (AIVCs)~\citep{bunne2024build,song2024toward}—offer a transformative, multiscale approach to simulating and analyzing molecules, cells, and tissues. The development of AIVCs offers a promising path forward by embedding molecular binding in a dynamic, data-integrated, and biologically realistic context. This systems-level perspective advances binding affinity prediction beyond isolated structural modeling, improving both its accuracy and translational relevance in computational drug discovery. We believe AIVCs can be leveraged to enhance binding affinity prediction in several key ways. \textit{1)~Contextual environment simulation.} Traditional models often assess binding in simplified, isolated systems. In contrast, AIVCs simulate full cellular environments, enabling context-aware predictions for complex targets, such as membrane proteins, and allosteric sites. \textit{2)~Dynamic protein conformations.} Binding is inherently dynamic. AIVCs can capture ligand entry and exit pathways, conformational transitions, and temporal fluctuations, providing a more realistic representation of binding processes. \textit{3)~Multi-target and off-target analysis.} By representing the entire proteome and interactome, AIVCs can evaluate binding specificity and off-target risks in parallel, improving early-stage selectivity assessments. \textit{4)~Integration of multi-omics and systems constraints.} AIVCs can incorporate transcriptomic, proteomic, metabolomic, and genomic data—enabling simulation of protein abundance, isoform specificity, metabolic competition, and mutation effects. This facilitates personalized affinity predictions and system-aware drug efficacy assessments by enabling cell type—or patient-specific predictions—key to advancing personalized medicine and precision pharmacology. 

Enabled by innovative dataset curation, novel algorithmic development, and creative evaluation, we envision an emerging paradigm shift in binding affinity prediction to advance precision pharmacology.

\subsubsection*{Acknowledgements}
We thank Nabil Faruk for constructive suggestions.
This work is supported in part by the RadBio-AI project (DE-AC02-06CH11357), U.S. Department of Energy Office of Science, Office of Biological and Environment Research, the Improve project under contract (75N91019F00134, 75N91019D00024, 89233218CNA000001, DE-AC02-06-CH11357, DE-AC52-07NA27344, DE-AC05-00OR22725), 
the Exascale Computing Project (17-SC-20-SC), a collaborative effort of the U.S. Department of Energy Office of Science and the National Nuclear Security Administration.

\bibliographystyle{plainnat}
\bibliography{reference}

\end{document}